**Time-domain Stokes mechanism of pair-correlated k-gap solitons in nonlinear photonic time-crystal slabs**

Jiaxiang Sun†, Guowei Chen†, Liang Zhang†, Yiming Pan*

State Key Laboratory of Quantum Functional Materials, School of Physical Science and Technology and Center for Transformative Science, ShanghaiTech University, Shanghai 200031, China

## Abstract

Pair generation in time-varying media is commonly attributed to time reflection at temporal boundaries or to amplification inside momentum $k$-gaps. Here we show that these two processes are connected by the "time-domain" Stokes phenomenon. A finite-duration photonic time-crystal (PTC) slab provides the necessary Stokes connection between the incident vacuum mode, transient k-gap amplification, and time-boundary scattering. With Kerr nonlinearity, the otherwise unbounded amplification is arrested, spawning Kerr-stabilized $k$-gap solitons. When these solitons cross the exit boundary of the time slab, Stokes-induced mode conversion produces a secondary pair-generation process, yielding four spatially separated and entangled pulse branches. Detection of a backward-propagating light pulse therefore heralds its forward-propagating partner. We further propose combined Hanbury-Brown–Twiss and Hong–Ou–Mandel measurements to test their nonclassical correlations. These results reveal a link between asymptotic Stokes physics and quantum temporal scattering in PTCs, and suggest a route toward ultrafast heralded quantum light sources.

Time-varying media have recently emerged as a versatile platform for exploring novel mechanisms such as photon pair generation, primarily through time reflection at abrupt temporal boundaries and parametric amplification within the momentum $k$-gap of photonic time crystals (PTCs) [1-6, 12-15]. Unlike spatial scattering, temporal scattering matches electromagnetic fields across a temporal interface while breaking energy conservation, enabling the active medium to transfer energy to the wave [3-5]. In the quantum regime, this process manifests as vacuum squeezing and generates entangled photon pairs [4,5,10,11]. Periodic temporal modulation further breaks continuous time-translation symmetry and opens a $k$-gap, where Floquet modes experience rapid exponential growth [12-15]. Since spatial homogeneity conserves momentum, a forward-propagating seed in a PTC necessarily couples to a counter-propagating partner via pair-generation processes [4,5,15].

However, direct observation of counter-propagating pairs in a linear PTC is intrinsically limited because modes inside the $k$-gap are frequency-locked and have zero group velocity [13-16]. Consequently, the amplified field remains spatially frozen. Kerr nonlinearity overcomes this constraint by saturating the unbounded $k$-gap amplification and reshaping the frozen modes into stable, propagating $k$-gap solitons [16]. Once formed, these Kerr-stabilized k-gap soliton branches can separate into spatially resolvable pulse pairs, providing a nonlinear route to observable pair generation. This deterministic pair generation shows that temporal boundaries and $k$-gaps are not passive scattering features, but active spatiotemporal regions that govern the creation, conversion, and routing of nonclassical light [5,6,15,16].

The abrupt mode conversions at temporal boundaries can be understood as a manifestation of the Stokes phenomenon [7,8]. Originally introduced in the asymptotic theory of Airy functions, this phenomenon describes the sudden "birth" of an exponentially small wave branch across a Stokes line [7-9]. It provides a general framework for wave scattering near turning points, including WKB tunneling in quantum mechanics [7-9] and Bragg reflection in spatial photonic crystals [30]. At a spatial bandgap edge, a propagating Bloch mode is converted into an evanescent mode, reflecting the same oscillatory-to-decaying transition found in the Airy turning-point problem. Temporal boundaries provide a distinct realization of this physics. Unlike spatial interfaces, which redirect wave energy, an abrupt temporal modulation breaks energy conservation and scatters waves in frequency and momentum space [3-5]. In the quantum regime, time reflection does not imply backward temporal propagation; rather, the modulation energy amplifies vacuum fluctuations to generate pairs of spatially counter-propagating photons. A seed pulse stimulates this process, producing a forward “signal” and a backward “idler” [4,5]. Thus, time reflection can be viewed as a temporal Stokes transition: the refractive-index jump acts as a temporal turning point that triggers a previously dormant solution branch [5,7-9]. Because causality enforces forward evolution in time, the generated "reflected" photon propagates forward temporally while carrying reversed spatial momentum.

Here we show that pair generation through time reflection and $k$-gap amplification is a consequence of a common mechanism: the time-domain Stokes phenomenon. We first map the wave equation within a finite-time PTC slab to a “soft”-boundary Airy model, showing that temporal-boundary scattering is governed by Stokes asymptotics. Because time evolution is causal, the resulting "reflection" appears as forward-time pair generation rather than backward-time propagation. We then show that Kerr nonlinearity overcomes the immobility of linear $k$-gap modes by saturating the parametric gain and transforming the frozen fields into propagating $k$-gap soliton branches [16]. When these solitons traverse the exit boundary of the modulation window, Stokes-induced splitting converts the localized field into spatially separated, entangled pulse pairs. This mechanism transforms temporal reflections into resolvable nonclassical pulse trains, linking classical wave asymptotics with quantum temporal scattering and suggesting a route toward heralded quantum light sources and ultrafast spatiotemporal control.

**Modelling.** To capture the temporal scattering and soliton dynamics, we model the finite-time nonlinear PTC slab within a one-dimensional electrodynamic framework. In time-varying media, it is convenient to formulate the dynamics in terms of the electric displacement field $D$ rather than the electric field $E$ [16,17], because Gauss's law, $\nabla \cdot D = 0$, enforces the continuity of $D$ across abrupt temporal boundaries and avoids singular field matching. We consider a centrosymmetric medium with a time-modulated linear permittivity and a Kerr nonlinearity. Assuming weak modulation and weak nonlinearity, the constitutive relations can be inverted and substituted into Maxwell's equations [16,17], yielding the nonlinear wave equation for the reduced displacement field: $\tilde{E} \equiv D$:

$$\frac{1}{c^2}\frac{\partial^2 \tilde{E}}{\partial t^2} = (1-\delta_1 cos\Omega t)\frac{\partial^2 \tilde{E}}{\partial x^2} - \beta\left|\tilde{E}\right|^2\frac{\partial^2 \tilde{E}}{\partial x^2}, \tag{1}$$

where $c$ is the background phase velocity, $\delta_1 \ll 1$ is the modulation depth, $\Omega$ is the driving frequency, and $\beta \ll 1$ is the effective Kerr coefficient. Using the plane-wave expansion $D = \widetilde{E_k}(t)e^{ikx}$, Eq. 1 reduces to a nonlinear Mathieu-type equation [12,14,16]

$$\frac{\partial^2}{\partial \tau^2}\tilde{E}_k + \left[\left(1+\tilde{d}\right) - \tilde{\delta}\cos 2\tau\right]\tilde{E}_k - \tilde{\beta}\left|\tilde{E}_k\right|^2\tilde{E}_k = 0 \tag{2}$$

This momentum-space equation provides the starting point for mapping PTC slab scattering to Stokes transitions.

**Stokes phenomenon of the PTC slab.** We first consider the linear PTC slab ( $\tilde{\beta} = 0$) to resolve mode connection across its temporal boundaries. In momentum space, the wave dynamics reduce to the Mathieu equation [12,14,15]. Near parametric resonance ($\tilde{d} \to 0$), the fast oscillation can be factored out, and the slowly varying envelope $\tilde{y}(\tau)$ of $\widetilde{E_k}$ obeys an inverted-oscillator equation

$$\frac{\partial^2 \tilde{y}}{\partial \tau^2} - \mu^2 \tilde{y} = 0. \tag{3}$$

The growth rate $\mu = \frac{1}{2}\sqrt{(\tilde{\delta}/2)^2 - \tilde{d}^2}$ is determined by the $k$-gap dispersion and controls the exponential amplification inside the unstable region. As the effective modulation depth $\tilde{\delta}$ varies across the finite switch-on/off boundaries of the slab, this envelope equation smoothly maps to the Airy equation near the temporal turning points [7-9], as derived in the Methods. This mapping identifies temporal-boundary mode conversion of PTC as a Stokes phenomenon.

To describe the switch-on dynamics, we approximate the soft temporal interface by a linear ramp of the effective modulation depth, $\tilde{\delta}(\tau) \approx \widetilde{\delta_0} + \gamma\tau$, where $\gamma$ is the ramp rate. Taking the bandgap edge at $\tau = 0$ gives $\left(\widetilde{\delta_0}/2\right)^2 = \widetilde{d^2}$. Over a short transition interval, the squared parametric growth rate becomes $\mu^2(\tau) = \left(\frac{\widetilde{\delta_0}+\gamma\tau}{2}\right)^2 - \widetilde{d^2} \approx \frac{\widetilde{\delta_0}\gamma}{2}\tau \equiv \eta\tau$, where $\eta > 0$. Substituting into Eq.3 yields $\frac{\partial^2 \tilde{y}}{\partial \tau^2} - \eta\tau\tilde{y} = 0$. With the rescaled coordinate $\zeta = \eta^{1/3}\tau$, the envelope equation reduces to the Airy form,

$$\frac{\partial^2 \tilde{y}}{\partial \zeta^2} - \zeta\tilde{y} = 0, \tag{4}$$

with solution $\tilde{y}(\zeta) = C_1\mathrm{Ai}(\zeta) + C_2\mathrm{Bi}(\zeta)$. For $\zeta > 0$, corresponding to the unstable PTC region, the two branches become exponentially decaying and growing,

$$\mathrm{Ai}(\zeta) \sim \frac{1}{2\sqrt{\pi}}\zeta^{-1/4}\exp\left(-\frac{2}{3}\zeta^{3/2}\right), \quad \mathrm{Bi}(\zeta) \sim \frac{1}{\sqrt{\pi}}\zeta^{-1/4}\exp\left(\frac{2}{3}\zeta^{3/2}\right). \tag{5}$$

The exponentially dominant Bi branch continuously connects to $k$-gap amplification. For $\zeta < 0$, corresponding to the stable unmodulated region, the Airy functions transition into oscillatory propagating modes, representing the incident and reflected waves:

$$\begin{aligned} Ai(\zeta) &\sim \frac{1}{\sqrt{\pi}}|\zeta|^{-1/4} sin\left(\frac{2}{3}|\zeta|^{3/2} + \frac{\pi}{4}\right), \\ \mathrm{Bi}(\zeta) &\sim \frac{1}{\sqrt{\pi}}|\zeta|^{-1/4}\cos\left(\frac{2}{3}|\zeta|^{3/2} + \frac{\pi}{4}\right). \end{aligned} \tag{6}$$

Note that the Stokes phase $\pi/4$ in the oscillatory tails is the asymptotic signature of temporal-boundary mode conversion [7-9] and fixes the phase relation between the incident and generated branches [7-9]. By symmetry, the same Airy mapping applies to the switch-off temporal boundary at $\tau = \tau_2$, as shown in Fig. 1a. Kerr nonlinearity saturates the $k$-gap amplification and reshapes the field into stable $k$-gap solitons [16], but it does not alter the local soft-boundary structure.

When the k-gap soliton branch reaches the exit boundary, it undergoes the same Stokes transition, enabling deterministic extraction of the pair-generation coefficients, as verified below by FDTD simulations. Thus, temporal scattering in a finite PTC slab can be described as a time-domain Stokes transition in both the linear and weakly nonlinear regimes.

Figure 1 depicts the finite-time PTC slab and illustrates pair generation as an asymptotic mode-conversion process. As Horsley and Pendry noted, in time-varying media "Maxwell's equations tell a quantum story": classical Maxwell dynamics can be reinterpreted as Bogoliubov mixing between positive- and negative-frequency modes, leading to photon-pair generation [29]. Our analysis identifies the time-domain Stokes phenomenon as the mathematical origin of the conversion. By relating the temporal boundaries [Figs. 1(a) and 1(c)] to the local Airy/Stokes branching process [7-9] at the exit interface [Fig. 1(b)], we demonstrate how a localized growing mode bifurcates into two oscillatory output branches across a temporal interface. This transition is the temporal analogue of the Airy turning-point connection between evanescent and oscillatory waves, indicating that time-reflected pair generation is a physical manifestation of Stokes-induced wave scattering [4,5,7-9].

In the complex plane, an abrupt temporal boundary acts as a Stokes line: crossing it activates a previously negligible wave branch and converts modulation energy into real photon pairs. Spatial momentum conservation places the generated photons in counter-propagating modes. Since time reflection involves Bogoliubov mode mixing in the quantum regime [4,5], the initial quantum vacuum is transformed into a two-mode squeezed state. Cosequently, the time-reflected and transmitted waves form an entangled pair with correlations fixed by the Bogoliubov coefficients. Bound by the irreversible arrow of time, this Stokes excitation is unidirectional: the generated branch propagates forward in time while carrying reversed spatial momentum.

**Secondary pair generation at temporal boundaries.** To test the Stokes-based mechanism and capture the nonlinear soliton dynamics, we perform FDTD simulations of a nonlinear PTC slab (Fig. 2). The system is seeded by a broadband 30-fs Gaussian pulse ( $\lambda_0 \approx 800$ nm) as the temporal modulation is switched on at $t = 0$. After a finite active duration of 100 fs, the modulation is switched off and the medium returns to its unmodulated dielectric state. This temporal window defines the PTC slab, enabling $k$-gap soliton formation and secondary pair generation at the exit boundary [12,16]. Numerical details are given in the Supplemental Material.

The spatiotemporal evolution shows that the finite-time PTC slab acts as a temporal-boundary converter for k-gap soliton branches (Fig. 2). After crossing the switch-on boundary, the weak seed pulse undergoes localized $k$-gap amplification. Before the field intensity diverges, Kerr nonlinearity arrests the exponential growth and restructures the field into counter-propagating $k$-gap solitons [16]. As these solitons traverse the switch-off boundary, a secondary Stokes-induced conversion splits them into four spatially resolvable output branches. Later, we show that this four-port emission provides a natural quantum detection scheme: the outer branches can be used for Hanbury Brown–Twiss (HBT) measurements, while the inner branches can be routed to Hong–Ou–Mandel (HOM) interferometry [20-22], as discussed below.

The resulting solitons regularize the parametric instability inherent to linear PTCs, where Floquet $k$-gap modes otherwise undergo unbounded exponential growth [13-15]. Although analogous to spatial Bragg solitons in nonlinear periodic structures [30], their kinematics are distinct. Bragg solitons form near frequency band edges and usually have small group velocities, whereas PTC $k$-gap solitons originate from Floquet band edges with a divergent dispersion slope, enabling the soliton envelope to undergo rapid envelope separation. This rapid envelope separation follows the causal forerunner analysis of [16]: The rapid spatial separation reflects reshaping of the wave packet centroid, while the information-carrying signal front (forerunner) remains bounded by the unmodulated speed of light. Thus, a nonlinear PTC provides a unique platform for generating and routing ultrafast k-gap soliton pulse envelopes with causal limits.

The stabilization of $k$-gap solitons relies on two effects of Kerr nonlinearity [16]. First, self-trapping compensates $k$-gap-induced dispersion. Second, the intensity-induced index shift detunes the local Floquet resonance from the $k$-gap center, thereby saturating the exponential gain. The field is then converted into a robust temporal pulse that propagates within the active PTC slab. At the exit boundary, the soliton encounters a temporal discontinuity that acts as a secondary Stokes turning point [5,7-9]. The resulting time-reflection process splits the half-harmonic ($\Omega/2$) component into a forward-transmitted pulse and a backward-propagating partner. Together with the pair generated at the switch-on interface, the finite-time slab emits four spatially distinct amplified pulses. Since each split pulse is associated with the Stokes mode conversion of vacuum fluctuations, these outputs should be viewed as entangled soliton-pair branches rather than classical wave fragments.

We further show that increasing the modulation strength changes the nonlinear balance of the PTC slab (Fig. 3). A stronger parametric drive reshapes the amplified field to condense into a multi-peaked, higher-order soliton train [16]. Upon reaching the exit boundary, secondary Stokes conversion acts on this structured envelope and produces a multiplexed sequence of output pulses. Since these pulses originate from the same vacuum-squeezing and Stokes processes, they are expected to retain nonclassical correlations across the emitted train. This amplitude-dependent response suggests that a finite-time PTC slab can generate tunable, high-repetition-rate entangled soliton trains for time-frequency quantum communication and ultrafast metrology.

**HBT and HOM verification.** To characterize the nonclassical statistics of the generated pulses, we quantize the relevant $k$-gap modes [5,18,19]. For a momentum pair $k = \pm k_0$ near the $k$-gap center, the effective Hamiltonian including Kerr nonlinearity reads

$$\hat{H}_k \ = \ \hbar\kappa\left(\hat{a}_k^\dagger \, \hat{a}_{-k}^\dagger + \hat{a}_k \, \hat{a}_{-k}\right) + \ \frac{\hbar U}{2}\left[\, \hat{n}_k^2 + \hat{n}_{-k}^2 + 4\hat{n}_k \, \hat{n}_{-k} - (\hat{n}_k + \hat{n}_{-k})\right], \tag{7}$$

The parametric term conserves the photon-number difference ( $Q = \hat{n}_k - \hat{n}_{-k}$), so a vacuum initial state evolves within the paired Fock subspace, $|\psi(t)\rangle = \sum_{n=0}^{\infty} c_n(t)|n,n\rangle$, where $n$ is the photon pair number. In the linear limit ($U \rightarrow 0$), the $k$-gap dynamics reduces to a degenerate optical parametric amplifier pumped at $2\omega_0$. Vacuum fluctuations are amplified into a two-mode squeezed state [4,5,18,19] with squeezing parameter $r = \kappa t = (\Delta n/2n_0)\omega_0 t$, see the derivation in Method M3 and M4. Assuming a Gaussian temporal coherence function $\gamma(\tau) = \exp(-\tau^2/2\tau_c^2)$, the delayed Hanbury Brown–Twiss (HBT) cross-correlation and Hong–Ou–Mandel (HOM) coincidence take the following form:

$$g_{HBT}^{(2)}(\tau) = 1 + \left(1 + \frac{1}{\sinh^2 r}\right) e^{-\tau^2/\tau_c^2} ,$$
$$g_{HOM}^{(2)}(\tau) = 1 - \left[1 - \left(\frac{\cosh r}{\cosh r + 1}\right)^2\right] e^{-\tau^2/\tau_c^2}. \tag{8}$$

These expressions predict a zero-delay HBT bunching peak and a HOM coincidence dip [20-22], providing direct signatures of pairwise quantum correlations. In a purely linear PTC, the photon-pair number grows exponentially with r ($\propto e^r$). The Kerr term ( $\hbar U$) detunes the resonance at high intensity, suppressing this divergence and stabilizing the macroscopic pair generation. The full quantum treatment of nonlinear k-gap soliton dynamics is discussed in a companion manuscript by Zhang et al. [34].

To verify the quantum nature of Stokes mode conversion in PTC slab, we map the four-port emission onto two interferometric measurements (Figs. 2c and 2d). The outer branches are sent to an HBT setup to probe temporal pair correlations. In the weak-pumping regime ($r = 0.3$), $g_{HBT}^{(2)}(\tau)$ exhibits a pronounced zero-delay peak, consistent with highly correlated signal-idler pairs in spontaneous parametric down conversion process [18,19]. This nonclassical feature weakens at higher gains ($r = 1.5$) due to the Kerr-induced suppression of multiphoton accidental coincidences. The inner branches are combined in an HOM configuration to test quantum indistinguishability. A coincidence dip appears at $\tau = 0$, reaching a normalized minimum of 0.25 for weak pumps, and becoming shallower as the gain increases. Together with a reference pulse for temporal calibration, these measurements provide an experimentally accessible protocol to verify both the nonclassical entanglement of the heralded $k$-gap solitons.

**Further discussions.**

Soliton splitting in a nonlinear PTC goes beyond linear wave packet scattering and can be viewed as a time-domain parametric process. The finite-time slab acts as an effective optical parametric oscillator [5,10,11,18,19]: temporal boundaries play the role of time mirrors [31], while the high-intensity $k$-gap soliton provides the pump that amplifies vacuum fluctuations into forward-signal and backward-idler pulse pairs. At Floquet resonance ($\Omega \simeq 2\omega_0$), this process is described by a Bogoliubov transformation and is closely related to the dynamical Casimir effect [10,11].

Although stimulated amplification may produce a dominant coherent field, the underlying two-mode squeezing and entanglement remain measurable through intensity-difference noise, HBT correlations, and HOM interference. Thus, the switch-on boundary and the soliton exit boundary operate as sequential quantum emitters, generating spatially separated and trackable entangled light pulses with femtosecond timing and broad tunability.

For experimental realization, we address that this Stokes-induced pair generation is intrinsically quantum optical and therefore requires optical-frequency quantum platforms that can access vacuum fluctuations, rather than being inferred solely from classical wave analogues such as water-wave time mirrors or microwave time-boundary experiments [31,32]. Our simulations indicate feasible operating conditions, including a modulation depth of $\Delta n/n_0 \approx 0.2$, femtosecond temporal switching, and weak Kerr nonlinearity. A possible near-infrared implementation could use femtosecond-pumped epsilon-near-zero (ENZ) media [26,27] or silicon-on-insulator waveguides [33] to generate ultrafast index modulation around the near-infrared carrier frequency. Time-modulated metasurfaces provide another route for realizing photonic time crystals and momentum-bandgap physics [28]. Combining such transient index change with a Kerr-active medium may enable experimental observation of k-gap soliton branches and their heralded entangled outputs

**Conclusion.** We have elucidated the temporal-boundary conversion of Kerr-stabilized k-gap soliton branches in a PTC slab. By mapping temporal boundaries to a time-domain Stokes transition, we show how periodic modulation amplifies Floquet modes, Kerr nonlinearity stabilizes them into propagating k-gap soliton branches, and the exit boundary converts them into spatially separated entangled photon pairs. HBT and HOM measurements are proposed to experimentally confirm this time-domain Stokes phenomenon and its nonclassical correlations. This framework connects classical wave asymptotics with quantum temporal scattering and identifies nonlinear PTCs as a platform for macroscopic entangled pulse generation.

Looking forward, this temporal-scattering paradigm suggests new opportunities for quantum photonics based on tunable entangled pulse trains. Nonlinear PTC slab may provide a route to high-dimensional quantum communication and ultrafast metrology, while the boundary conversion of k-gap soliton branches could enable fast temporal switching and optical gating. More broadly, these results show how time-varying media can be used to control both classical wave dynamics and macroscopic quantum correlations.

## Acknowledgments

Y. P. acknowledges the support of the NSFC (No. 2023X0201-417-03) and the start-up fund from ShanghaiTech University.

Correspondence and requests for materials should be addressed to Y.P.: yiming.pan@shanghaitech.edu.cn.

## Data and code availability

The code repository will be made **publicly** available on acceptance through an open-source link.

## Conflict of interest

The authors declare no competing interests.

## Methods

### M1. Stokes phenomena in the Airy equation

The Airy equation provides the canonical local model for wave dynamics near a turning point [7-9]. It connects an oscillatory propagating regime to an exponential growing/decaying regime, which is the basic asymptotic structure underlying the Stokes phenomenon:

$$\frac{d^2y}{d\zeta^2} - \zeta y = 0. \qquad (M1)$$

The sign of $\zeta$ separates the two regimes. For $\zeta < 0$, the Airy solutions are oscillatory and represent propagating modes on the unmodulated side of the temporal boundary. For $\zeta > 0$, they become exponential branches, corresponding to the growing and decaying modes inside the PTC $k$-gap.

In the oscillatory limit, $\zeta \to -\infty$,

$$\begin{aligned} \mathrm{Ai}(\zeta) &\sim \frac{1}{\sqrt{\pi}(-\zeta)^{1/4}} \sin\left[\frac{2}{3}(-\zeta)^{3/2} + \frac{\pi}{4}\right] \\ \mathrm{Bi}(\zeta) &\sim \frac{1}{\sqrt{\pi}(-\zeta)^{1/4}} \cos\left[\frac{2}{3}(-\zeta)^{3/2} + \frac{\pi}{4}\right] \end{aligned}, \qquad (M2)$$

These oscillatory asymptotics are identified as the transmitted and time-reflected branches. The time-reflected branch does not propagate backward in time; but is a forward-evolving conjugate mode with reversed spatial momentum.

In the exponential limit, $\zeta \to +\infty$,

$$\begin{aligned} \mathrm{Ai}(\zeta) &\sim \frac{1}{2\sqrt{\pi}\zeta^{1/4}} \exp\left[-\frac{2}{3}\zeta^{3/2}\right] \\ \mathrm{Bi}(\zeta) &\sim \frac{1}{\sqrt{\pi}\zeta^{1/4}} \exp\left[\frac{2}{3}\zeta^{3/2}\right] \end{aligned}, \qquad (M3)$$

Here, $\mathrm{Ai}(\zeta)$ represents the decaying branch, while $\mathrm{Bi}(\zeta)$ represents the growing branch, which is the local analogue of the $k$-gap-amplified mode in the PTC slab. Analytic continuation across the turning point converts this exponential branch into oscillatory output branches [7-9], corresponding to the Stokes activation of a previously subdominant component.

In a finite-time PTC slab, the switch-on and switch-off boundaries acct as two temporal turning points. The first connects the incident seed or vacuum mode to the $k$-gap branches, while the second converts the amplified branch back into propagating output modes. This Airy connection underlies our interpretation of time reflection and secondary pair generation as the same temporal-boundary Stokes process.

### M2. FDTD numerical simulation of the nonlinear finite-time PTC slab

We simulate the nonlinear finite-time PTC slab using a scalar field $u(x,t)$ , governed by [16,17]

$$\frac{1}{c^2}\frac{\partial^2}{\partial t^2}\left[\epsilon(x,t)u + \chi^{(3)}|u|^2 u\right] = \frac{\partial^2 u}{\partial x^2}, \qquad (M4)$$

where $\epsilon(x,t)$ is the spacetime-modulated dielectric function, $\chi^{(3)}$ is the Kerr coefficient, and $c$ is the background light speed. In the simulations, the modulation is spatially uniform, $\epsilon(t,x) = \epsilon(t)$ , and normalized units with $c = 1$ are used.

Under weak-modulation and weak-nonlinearity approximations, Eq. (4) reduces to the displacement-field equation used for the analytical discussion,

$$\frac{1}{c^2}\frac{\partial^2 u}{\partial t^2} = [1 - \delta_1 \cos(\Omega t)]\frac{\partial^2 u}{\partial x^2} - \beta|u|^2\frac{\partial^2 u}{\partial x^2}, \qquad (M5)$$

where $\delta_1$ is the modulation depth , $\Omega$ is the modulation frequency, and $\beta$ is the nonlinear coefficient. Equation (M5) is used to interpret the Floquet resonance, $k$-gap opening, and Stokes mapping, while the FDTD results are obtained from Eq. (M4).

The finite-time PTC slab is implemented through a hyperbolic-tangent modulation profile,

$$\epsilon(t) = 1 + \frac{\chi_1}{2}\left[\tanh\big(\alpha(t - t_1)\big) - \tanh\big(\alpha(t - t_2)\big)\right]\cos(\Omega t), \qquad (M6)$$

Here, $\chi_1$ sets the modulation amplitude, $t_1$ and $t_2$ define the switch-on and switch-off times, and $\alpha$ controls boundary sharpness. This profile separates the evolution into an initial unmodulated region, an active PTC region, and a final unmodulated region.

The incident field is initialized as a Gaussian-enveloped carrier wave,

$$u(x,0) = A_0 \exp(ik_0 x)\exp\left[-\frac{(x - x_c)^2}{2\sigma^2}\right], \qquad (M7)$$

where $A_0$, $\sigma$ and $x_c$ are the seed amplitude, envelope width, and initial pulse center. The carrier is chosen near the half-harmonic resonance, $\omega_0 = \Omega/2$, so that the active modulation opens a momentum $k$-gap. During the PTC window, the field is amplified through the growing Floquet branch and is converted at the switch-off boundary into transmitted and time-reflected output components.

The spatial and temporal domains are discretized on uniform grids. The spatial derivative is evaluated by a central-difference scheme, while the nonlinear temporal evolution is treated semi-

implicitly. Defining

$$A_n = \epsilon_n + \chi^{(3)}|u_n|^2, \tag{M8}$$

the update equation is

$$\left[\epsilon_{n+1} + \chi^{(3)}|u_{n+1}|^2\right]u_{n+1} = 2A_n u_n - A_{n-1}u_{n-1} + \Delta t^2(\partial_x^2 u)_n \tag{M9}$$

This local nonlinear update is reduced to a scalar cubic equation for $r = | u_{n+1} |$,

$$\chi^{(3)} r^3 + \epsilon_{n+1} r = |R_n|. \tag{M10}$$

where $R_n$ denotes the right-hand side of Eq. (9). The phase of $u_{n+1}$ is taken from $R_n$, and the cubic equation is solved by Newton-Raphson iteration initialized from the corresponding linear response. This semi-implicit scheme provides stable treatment of Kerr feedback and allows direct observation of seed amplification, nonlinear gain arrest, $k$-gap soliton formation, and secondary Stokes splitting at the exit boundary.

**M3. Quantization and squeezing of linear PTC**

We first consider the linear quantum limit of the PTC to identify the pair-generation channel associated with the momentum $k$-gap [4,5,15]. The refractive index is harmonically modulated as

$$n(t) = n_0 + n_1 \cos(\Omega t), \qquad n_1 \ll n_0, \tag{M11}$$

with relative modulation depth $\delta = n_1/n_0$. For a momentum mode $k$, the unperturbed frequency is $\omega_0 = ck/n_0$. To first order in $\delta$, the modulation induces a resonant two-photon interaction between opposite-momentum modes $k$ and $-k$,

$$\hat{H}_{\text{int}}(t) = \hbar\eta\cos(\Omega t)\left(\hat{a}_k^\dagger \hat{a}_{-k}^\dagger + \hat{a}_k \hat{a}_{-k}\right), \tag{M12}$$

where $\eta \propto \delta\omega_0$ is the effective parametric coupling. Thus, temporal modulation at $\Omega$ can resonantly create photon pairs in two opposite-momentum modes.

Moving to the rotating frame and applying the rotating-wave approximation, we define the detuning $\Delta = \Omega - 2\omega_0$ and the resonant coupling $g = \eta/2$. Bogoliubov growth rate is[4,5,18,19]

$$\Gamma = \sqrt{g^2 - \left(\frac{\Delta}{2}\right)^2}. \tag{M13}$$

Exponential amplification occurs for $| \Delta | < 2g$, which defines the quantum $k$-gap. At the center of the gap, $\Delta = 0$, the slowly varying operators obey the Bogoliubov transformation

$$\hat{a}_k(t) = u(t)\hat{a}_k(0) + v(t)\hat{a}_{-k}^{\dagger}(0), \tag{M14}$$

with

$$u(t) = \cosh r\,, \qquad v(t) = -i\sinh r\,, \qquad r = gt. \tag{M15}$$

The transformation is canonical because

$$|u(t)|^2 - |v(t)|^2 = 1. \tag{M16}$$

Therefore, $k$-gap amplification in a linear PTC is a Bogoliubov process rather than ordinary scalar gain: it drives the initial vacuum state to a two-mode squeezed state,

$$|\psi(t)\rangle = \frac{1}{\cosh r}\sum_{n=0}^{\infty}\left[-e^{i\phi}\tanh r\right]^{n}|n_k, n_{-k}\rangle\,. \tag{M17}$$

The mean photon number in each mode is $N(t) = \sinh^2 r$. Thus, a linear PTC $k$-gap acts as a time-domain optical parametric amplifier.

To include the leading nonlinear correction, we use an effective two-mode Hamiltonian [16,18],

$$\widehat{H}_{\text{eff}} = \hbar\kappa\left(\hat{a}_k^{\dagger}\hat{a}_{-k}^{\dagger} + \hat{a}_k\hat{a}_{-k}\right) + \frac{\hbar U}{2}\left(\hat{n}_k^2 + \hat{n}_{-k}^2 + 4\hat{n}_k\hat{n}_{-k} - \hat{n}_k - \hat{n}_{-k}\right), \tag{M18}$$

where $\kappa$ is the resonant pair-generation rate and $U$ characterizes the Kerr-induced nonlinear potential. The pair-generation term conserves the photon-number difference $\hat{Q} = \hat{n}_k - \hat{n}_{-k}$, so a state initialized from vacuum remains in the paired Fock subspace. In the linear limit $U \to 0$, the corresponding coefficients reduce to the two-mode squeezed distribution above. For finite $U$, they are determined by the nonlinear Hamiltonian evolution.

**M4. HBT and HOM second-order correlation functions**

We use second-order correlation measurements to characterize the pair correlation and indistinguishability of the generated signal-idler pulses. The signal mode $s$ and idler mode $i$ are identified with the two opposite-momentum output branches generated by the temporal boundary or by secondary soliton splitting.

For the Hanbury Brown–Twiss measurement, the signal-idler cross-correlation [20,21],

$$g_{s,i}^{(2)}(0) = \frac{\langle\hat{n}_s\hat{n}_i\rangle}{\langle\hat{n}_s\rangle\langle\hat{n}_i\rangle}, \tag{M19}$$

Using the two-mode squeezed state derived in M3 and assuming a Gaussian temporal coherence envelope, the delayed HBT cross-correlation is

$$g^{(2)}_{\mathrm{HBT}}(\tau) = 1 + \left(1 + \frac{1}{\sinh^2 r}\right) \exp\left[-\frac{\tau^2}{\tau_c^2}\right], \qquad (M20)$$

This function approaches unity at large delay and exhibits a zero-delay bunching peak, reflecting pair correlation between signal and idler photons. As $r$ increases, multiphoton accidental coincidences reduce the relative contrast of the peak.

The Hong–Ou–Mandel measurement probes the indistinguishability of the paired photons [22]. For the full two-mode squeezed input, multiphoton components reduce the unheralded HOM visibility. With threshold detectors, the visibility is

$$V(r) = 1 - \left[\frac{\cosh r}{\cosh r + 1}\right]^2, \qquad (M21)$$

It approaches 75% in the weak-squeezing limit and vanishes in the strong-squeezing limit, where multiphoton accidental coincidences dominate. Including the same Gaussian temporal coherence envelope, the normalized HOM coincidence function is

$$g^{(2)}_{\mathrm{HOM}}(\tau) = 1 - V(r) \exp\left[-\frac{\tau^2}{\tau_c^2}\right], \qquad (M22)$$

This function has a unit background at large delay and forms a coincidence dip at $\tau = 0$. Weak squeezing suppresses multiphoton contamination and yields a deeper dip, whereas stronger squeezing increases accidental coincidences and reduces the HOM visibility. For the nonlinear slab, the same observables are evaluated using the paired state obtained from the nonlinear Hamiltonian in M3. The analytical expressions above therefore serve as the linear reference limit. Deviations from these curves quantify the effects of Kerr-induced detuning, gain arrest, and macroscopic $k$-gap soliton occupation.

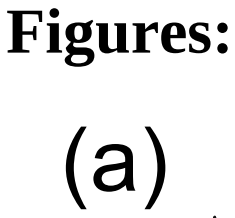

**Figures:**

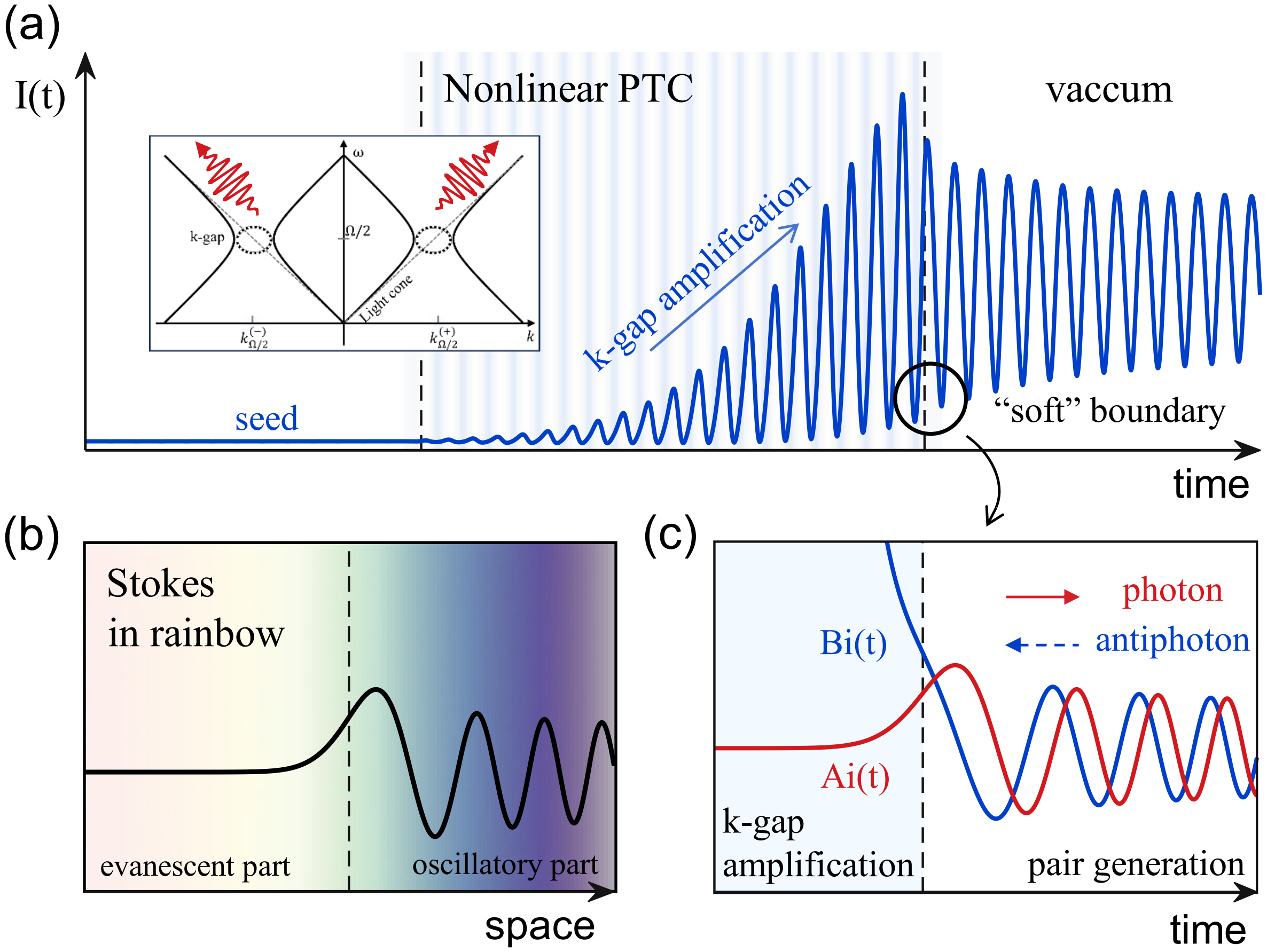

**FIG. 1. The time-domain Stokes phenomenon and photon pair generation in a finite-time photonic time-crystal (PTC) slab. (a)** Macroscopic temporal scattering dynamics. An incident seed pulse undergoes exponential amplification within the PTC $k$-gap (inset) and subsequently scatters at the exit boundary into oscillatory output modes. **(b)** Spatial Stokes analogy, illustrating the classical transition from an evanescent to an oscillatory regime across a turning point. **(c)** Microscopic Airy/Stokes mode conversion at the second temporal boundary. The localized growing branch $[B_i(t)]$ abruptly bifurcates into two counter-propagating oscillatory states (photon and antiphoton), physically dictating the time-reflected pair generation. The Airy/Stokes interpretation follows the standard turning-point connection theory [7–9].

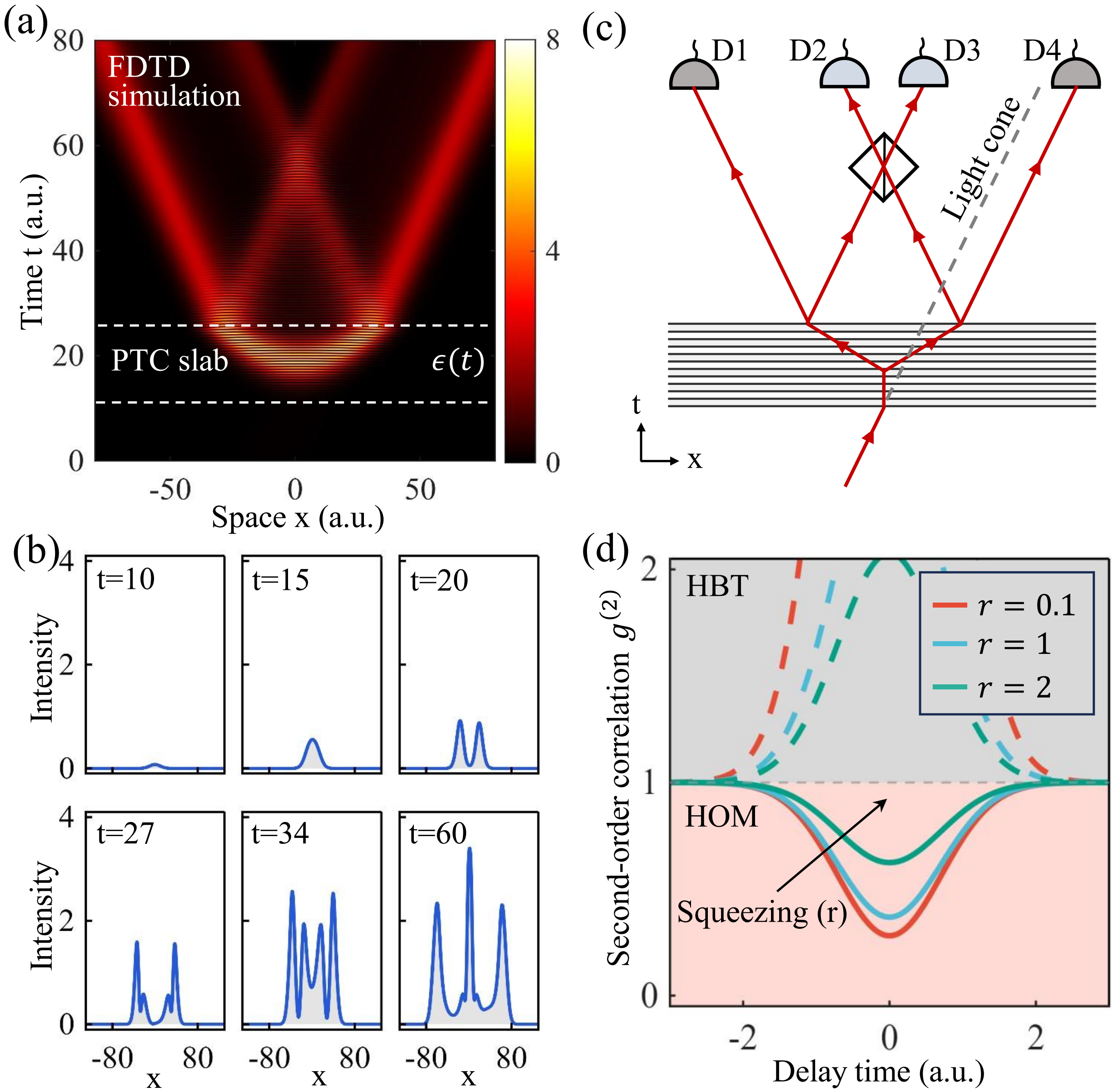

**FIG. 2. Spatiotemporal dynamics and quantum statistical verification of heralded k-gap soliton pairs. (a),(b)** Full-wave FDTD simulation (a) and corresponding intensity snapshots (b) of a finite-time nonlinear PTC slab acting as a superluminal pulse propagator. An incident seed undergoes localized amplification, reorganizes into a counter-propagating soliton pair, and experiences a secondary Stokes splitting at the exit boundary to emit four spatially resolvable branches. **(c)** Proposed four-port measurement topology. The outermost branches are designated for Hanbury Brown–Twiss (HBT) cross-correlation, while the inner branches are routed for Hong–Ou–Mandel (HOM) interferometry. The proposed correlation measurements follow the standard HBT and HOM schemes [20–22]. **(d)** Analytical second-order correlation functions $g^{(2)}(T)$. The nonclassical bipartite entanglement is rigorously verified by the drastic transition from a strong bunching peak $g^{(2)} \gg 2$ in the HBT scheme to a perfect anti-bunching dip ($g^{(2)} < 1$) in the HOM scheme. Lowering the PTC modulation strength (smaller squeezing parameter $r$) effectively suppresses multiphoton emissions, thereby maximizing the indistinguishability of the entangled soliton pairs.

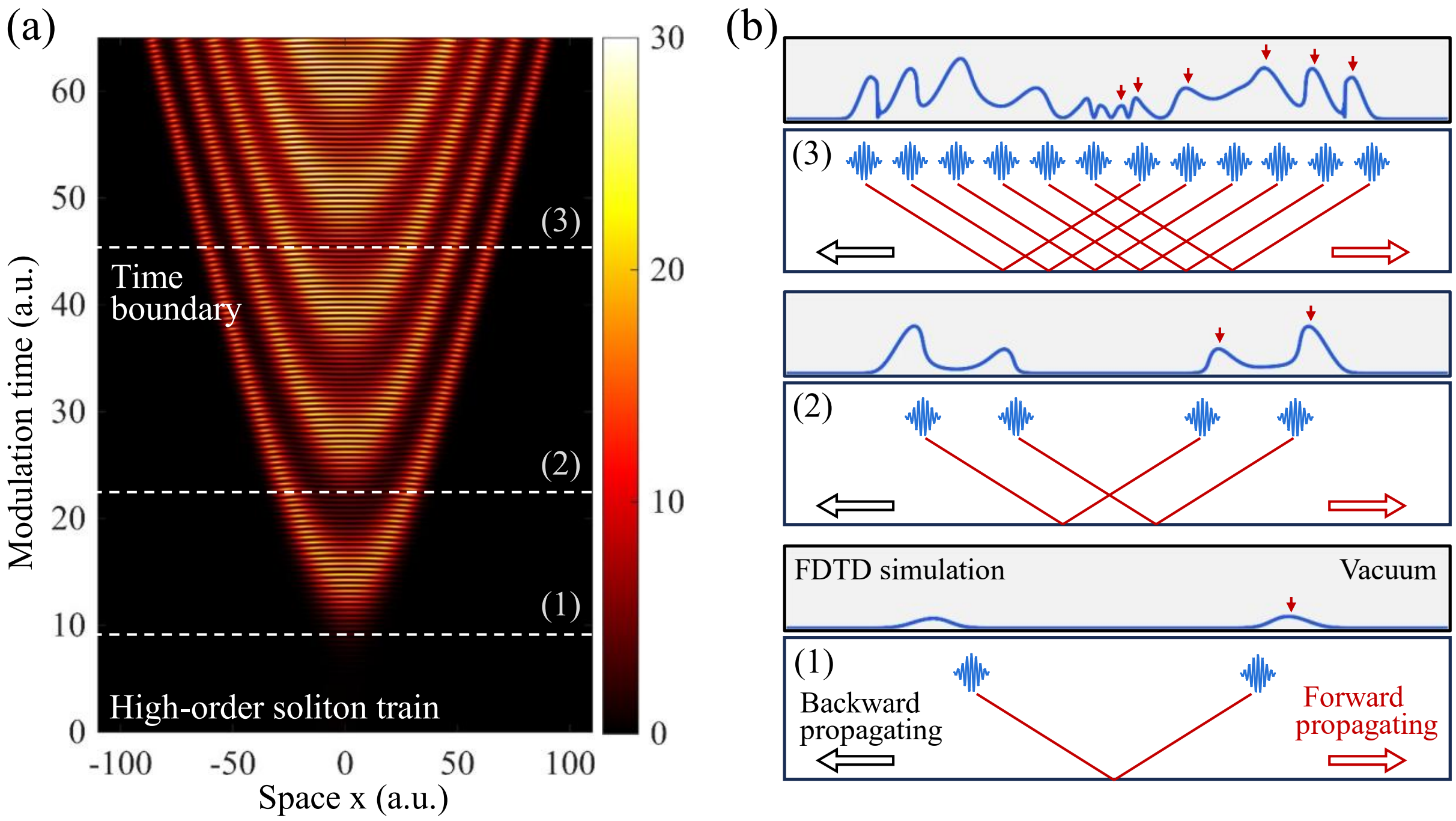

**FIG. 3.** Generation of entangled light pulse trains via higher-order $k$-gap solitons. The spatiotemporal dynamics (left) and output schematics (right) illustrate the dependence of soliton splitting on the PTC modulation depth. **(1)** Weak modulation limits the system to fundamental $k$-gap amplification, emitting a single pair of light pulses. **(2)** Moderate modulation excites second-order k-gap soliton dynamics, cascading the boundary scattering into two distinct pairs of pulse trains. **(3)** Strong modulation drives the formation of higher-order $k$-gap solitons, producing a dense, multiplexed cascade of entangled pulses. This amplitude-dependent scaling provides an on-demand mechanism for generating high-repetition-rate entangled pulse trains.